# Effects of Al content on the oxygen permeability through dual-phase membrane $60Ce_{0.9}Pr_{0.1}O_{2-\delta}$-$40Pr_{0.6}Sr_{0.4}Fe_{1-x}Al_xO_{3-\delta}$

Lei Shi[a], Shu Wang[a], Tianni Lu[b], Yuan He[a], Dong Yan[a], Qi Lan[a], Zhiang Xie[a], Haoqi Wang[a], Mebrouka Boubeche[a], Huixia Luo[a]*

[a]School of Material Science and Engineering and Key Lab Polymer Composite & Functional Materials, Sun Yat-Sen University, No. 135, Xingang Xi Road, Guangzhou, 510275, P. R. China

[b]School of Materials Sciences and Engineering, Shenyang Aerospace Unversity, Shenyang, 110136, P. R. China

*Corresponding author/authors complete details (Telephone; E-mail:) (+0086)-2039386124

luohx7@mail.sysu.edu.cn

Abstract graphic:

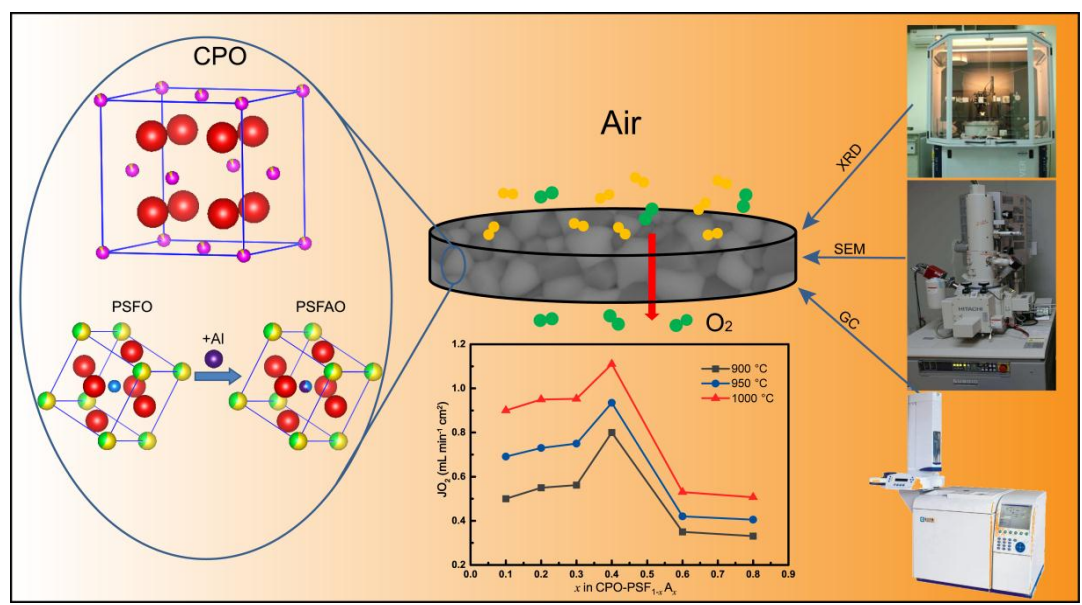

**ABSTRACT:**

Ceramic dual-phase oxygen transport membranes with the composition of 60wt.% $Ce_{0.9}Pr_{0.1}O_{2-\delta}$-40wt.%$Pr_{0.6}Sr_{0.4}Fe_{1-x}Al_xO_{3-\delta}$ ($x$ = 0.05, 0.1, 0.2, 0.3, 0.4, 0.6, 0.8, 1.0) (60CPO-40PSF$_{1-x}$A$_x$O) based on $60Ce_{0.9}Pr_{0.1}O_{2-\delta}$-$40Pr_{0.6}Sr_{0.4}FeO_{3-\delta}$ doped Al was successfully synthesized through a modified Pechini method. Crystal structure, surface microtopography and oxygen permeability are investigated systematically. The cell parameters of perovskite phase first increased and then decreased with the increase of Al content, which is related to the radius of the $Al^{3+}$ and the formation of impurity phase. As $x$ ranges from 0.1 to 0.8, the oxygen permeability of the materials first increases and then decreases, and the maximum value of oxygen permeation rate for 60CPO-40PSF$_{1-x}$A$_x$O membranes with 0.4mm thickness at 1000 °C is 1.12 mL min$^{-1}$ cm$^{-2}$ when $x$ = 0.4. XRD measurements revealed high temperature stability and $CO_2$-tolerant property of the dual-phase composites. The partial replacement of $Fe^{3+}/Fe^{4+}$ by $Al^{3+}$ causes the material not only to exhibit good stability, but also to increase the oxygen permeability of the membranes.

**Introduction**

Ceramic composites containing mixed ionic-electronic conductivity (MIEC) have received extensive interests due to their attractive prospects in high-temperature electrochemical applications, such as cathodes of solid oxide fuel cells (SOFC), dense inorganic membranes for pure oxygen preparation, and chemical membrane reactors.[1-8] In addition, since the potential application of MIEC oxygen transport membranes (OTMs) in oxy-fuel combustion can significantly reduce $CO_2$ and toxic $NO_x$ emissions in power stations, the environmentally friendly ceramic membrane technology has become a new hotspot for researchers.[9-11] However, this technology puts forward two basic requirements for ceramic membrane materials: (1) it must possess as high as possible oxygen permeability; (2) it must have an excellent stability in a wide range of oxygen concentration and temperature.[12,13]

It has been previously established that single perovskite OTMs (*e.g.* $Ba_{0.5}Sr_{0.5}Co_{0.8}Fe_{0.2}O_{3-\delta}$, $La_{0.6}Sr_{0.4}Co_{0.2}Fe_{0.8}O_{3-\delta}$) have high oxygen permeability, but exhibit limited stability under large oxygen chemical potential gradients.[14-18] To improve stability of single perovskite MIEC OTMs, the researchers replaced the Co of perovskite in the *B*-site with metal cations with more stable oxidation state such as Al, Zr, and Ti, etc.[19-21] As an economical and environmentally friendly material, aluminum is especially widely used in the *B*-site doped perovskite oxides. For example, Holc et al. studied $La_{0.8}Sr_{0.2}Fe_{1-x}Al_xO_{3-\delta}$ and suggested that $La_{0.8}Sr_{0.2}Fe_{0.7}Al_{0.3}O_{3-\delta}$ (LSFA) could be a candidate for intermediate-temperature SOFC in view of LSFA's excellent stability.[20] Jiang et al. reported a novel cobalt-free $BaFe_{0.9}Zr_{0.05}Al_{0.05}O_{3-\delta}$ OTM reactor coupled with nitrous oxide decomposition methane and carbon dioxide reforming.[22] It has become an indisputable fact that the doping of Al can improve the structure stability of the perovskite OTMs, but the effect of Al doping on the oxygen permeability of the OTMs is still controversial. Some studies have found that the doping of Al significantly decreased the oxygen permeability, which can be attributed to the fact that the substitution of $Al^{3+}$ for $Co^{2+}$ and $Fe^{3+}/Fe^{4+}$ (e.g. $La_{1-y}Ca_yFe_{1-x}Al_xO_{3-\delta}$, $LaMg_{0.1}Ga_{0.9-x}Al_xO_{3-\delta}$, $SrCo_{0.4}Fe_{0.6-x}Al_xO_{3-\delta}$, $Sr_{0.7}La_{0.3}Fe_{1-x}Al_xO_{3-\delta}$ and

$Sr_{0.7}La_{0.3}Co_{0.8}Al_{0.2}O_{3-\delta}$) not only decreased the oxygen vacancies but also the conductivity of perovskite.[23-28] However, some researchers also found that the substitution of $Al^{3+}$ for $Co^{2+}$ and $Fe^{3+}/Fe^{4+}$ is not only enhance the stability but also the oxygen permeability. For instance, Julia et al. reported that the oxygen permeability of $Ba_{0.5}Sr_{0.5}Fe_{1-x}Al_xO_{3-\delta}$ ($0 \leq x \leq 0.2$) membranes has been improved with rising aluminum content from 0 to 0.1.[23] E. Babakhani et al. also found that the introduction of Al cation increased the oxygen permeability of BSCFO oxides.[29] Recently, Kaveh et al. used a small amount of Al doping into the *B*-site of the perovskite phase in the $60Ce_{0.9}Nd_{0.1}O_{2-\delta}$-$40Nd_{0.6}Sr_{0.4}FeO_{3-\delta}$ (CNO-NSFO) and found that 60wt.%$Ce_{0.9}Nd_{0.1}O_{2-\delta}$-40wt.%$Nd_{0.6}Sr_{0.4}Fe_{0.8}Al_{0.2}O_{3-\delta}$ exhibited higher oxygen permeability and much more stability in comparison to CNO-NSFO.[30,31] Zhu group is doped with Al on the basis of $Ce_{0.85}Sm_{0.15}O_{1.925}$-$Sm_{0.6}Sr_{0.4}FeO_{3-\delta}$ (SDC-SSFO), and the obtained $Ce_{0.85}Sm_{0.15}O_{1.925}$-$Sm_{0.6}Sr_{0.4}Al_{0.3}Fe_{0.7}O_{3-\delta}$ exhibits high oxygen permeability and stability and can work stably for a long time in the membrane reactor for the partial oxidation of methane.[32-34] Although many Al-containing MIEC membranes have been reported, the influence of the change of Al doping content in *B*-site of perovskite phase on the oxygen permeability and stability of the composite OTMs has been rarely studied so far. Here we report a series of Al-containing ceramic composite OTMs with the composition of 60wt.%$Ce_{0.9}Pr_{0.1}O_{2-\delta}$-40wt.%$40Pr_{0.6}Sr_{0.4}Fe_{1-x}Al_xO_{3-\delta}$, ($x$ = 0.1, 0.2, 0.3, 0.4, 0.6, 0.8, 1.0; 60CPO-40PSF$_{1-x}$A$_x$O). The purpose of this work is to understand the influence of the substitution of Al into Fe in the *B*-site of $Pr_{0.6}Sr_{0.4}FeO_{3-\delta}$ phase in our previous reported $60Ce_{0.9}Pr_{0.1}O_{2-\delta}$-$40Pr_{0.6}Sr_{0.4}FeO_{3-\delta}$ materials on phase structure transition, oxygen permeability and phase stability.[35]

**Experiment**

**Synthesis of powders and membranes**

The 60CPO-40PSF$_{1-x}$A$_x$O powders are prepared by a modified Pechini method. As shown in the **Fig. S1**, the corresponding nitrates were sequentially weighed and added into the deionized water according to the stoichiometric ratio of the material. Following, the citric acid monohydrate ($C_6H_8O_7 \cdot H_2O$) as a complexing agent and ethylene glycol

$(CH_2OH)_2$ as a dispersing agent were added to the solution. The obtained solution was heated and stirred using a magnetic stirrer to evaporate water to obtain a gel. Then the gel was dried in an oven at 150 °C and ground to obtain the powder precursor. The as-obtained precursor was heated at 600 °C to get rid of the organic components, and then calcined in a furnace at 950 °C for 10 h to obtain the powders of 60CPO-40PSF$_{1-x}$A$_x$O. The obtained powders were pressed to ~ 9.5 MPa using a cylindrical stainless steel mold with a radius of 7.5 millimeter to obtain green disks. The black disks are sintered at 1450 °C in air for 5 h with a heating/cooling rate of 1.5 °C/min to obtain the dense 60CPO-40PSF$_{1-x}$A$_x$O OTMs.

**Characterization of membranes**

The phase structures of the 60CPO-40PSF$_{1-x}$A$_x$O powders and membranes were judged by X-ray diffraction (XRD, D-MAX 2200 VPC, Rigaku with Cu Kα). The data set were recorded in a step-scan mode in the 2θ range of 20°-80° with an interval of 0.02°. The Rietveld's refinement of XRD data were analyzed by the Fullprof suite (version: 14-June-2018) software. The crystal structures were made by the Vesta program. The microstructures of the as-sintered membranes were observed by scanning electron microscopy (SEM, Quanta 400F, Oxford), backscattered electron microscopy (BSEM) and energy dispersive X-ray spectroscopy (EDXS).

**Oxygen permeability of membranes**

Oxygen permeability of 60CPO-40PSF$_{1-x}$A$_x$O composite membranes were evaluated by a homemade high-temperature oxygen transmission equipment.[36,37] The 60CPO-40PSF$_x$A$_{1-x}$O composite membranes was sealed on a corundum tube with a heat-resistant adhesive (Huitian, China) and baked at 140 °C for 10 h, and the lateral direction of the oxygen permeable membrane was also covered with heat-resistant adhesive to avoid the transmission of radial oxygen affecting the final measured value. The effective working areas of our membranes are around 0.8659 cm$^2$. One side of the composite membranes was feed by dry synthetic air which comprising of high purity $O_2$ and $N_2$ with ratio of 21:79. And He or $CO_2$ as a sweeping gas were fed to the other side of the membranes. All inlet gas flows are controlled by the mass flowmeters (Sevenstar, China) and are periodically calibrated using a soap membrane flow meter. Dry synthetic air (21 % $O_2$ + 79 % $N_2$) with a flow rate of 100 mL min$^{-1}$ as feed gas; a mixture of He or $CO_2$ (49 mL min$^{-1}$) and Ne (1 mL min$^{-1}$) of internal standard gas as sweeping gas. Analysis of the composites of the gas phase mixture using the gas chromatograph (GC, Agilent-7890B, USA). The oxygen permeation rate of the membrane were calculated by the following formula .

$$J_{O2}(mL\bullet cm^{-2}\bullet min^{-1}) = (C_{O2} - \frac{C_{N2}}{4.02}) \times \frac{F}{S} \quad (1)$$

Where $C_{O2}$ and $C_{N2}$ represent the oxygen concentration and the nitrogen concentration, respectively and they can be measured by gas chromatograph. 4.02 is the ratio of the leaked nitrogen according to the theory of Kundsen diffusion. There is a small amount of air that diffuses through the high temperature resistant adhesive, which will interfere with our judgment of the true oxygen permeation rate, so we must deduct the oxygen (< 7 %) that is not diffused through the membrane. $F$ is the total flow rate of the exhaust gas of the sweep side. It can be calculated based on the concentration of helium. And $S$ is the effective oxygen permeation working region of the 60CPO-40PSF$_{1-x}$A$_x$O dual-phase membrane sealed on the corundum tube.[38-40]

**Result and Discussion**

The crystal structures of the as-prepared 60CPO-40PSF$_{1-x}$A$_x$O powders after calcined at 950 °C for 10 h are determined by XRD. As shown in **Fig. 1**, the XRD patterns are mainly indexed to the combination of fluorite phase CPO (space group *No.*255: *Fm$\bar{3}$m*) and perovskite phase PSF$_{1-x}$A$_x$O (space group *No.*74: *Imma*). And it can be observed that the peak of PSF$_{1-x}$A$_x$O phase shifts to a higher Angle with the increase of $x$. In other words, as more Fe is replaced by Al, the cell parameters of PSFAO phase become smaller, which is attributed to the smaller ion radius of Al$^{3+}$ (50 pm) in comparison to Fe$^{3+}$ (64 pm). With the continuous increase of $x$, the peak position of the CPO phase did not change at all, while the peak of the PSF$_{1-x}$A$_x$O phase shifted continuously, indicating that Al was successfully added into the perovskite phase but not into the CPO phase. When $x \geq 0.6$, part of the XRD peaks of $PSF_{1-x}A_xO$ overlapped with part of the XRD peaks of CPO phases, which made the peak (32.5°, 46.9°, 58.6°, 78.7°) become wider. And when $x \geq 0.8$, the impurity phase $PrSrAl_3O_7$ appears in the XRD patterns. The above analysis indicated that the maximum solid solution Al in PSFO is 0.6. When $x \leq 0.6$, the composite powders consist of only CPO and PSFAO phases, suggesting that the 60CPO-40PSF$_{1-x}$A$_x$O ($0 \leq x \leq 0.6$) samples can be successfully synthesized via the modified Pechini method. In order to check phase structures of the 60CPO-40PSF$_{1-x}$A$_x$O composite membranes, the sintered membranes are also characterized by XRD (see **Fig. S2**). The results reveal that when $x \leq 0.6$, the sintered composite membranes under studied also consist of only CPO and $PSF_{1-x}A_xO$ phases, whereas if $x \geq 0.8$, the additional phase $PrSrAl_3O_7$ shows again in the XRD patterns for sintered samples, which is consistent with the XRD results of powder samples.

To further verify the above conclusions, we refined all the XRD patterns of the dual-phase powders using Rietveld refinement and obtained their cell parameters. The results showed that with the increase of aluminum content, the cell parameters of $PSF_{1-x}A_xO$ phase increased gradually, while the cell parameters of CPO phase did not change significantly, as shown in **Fig. 2**. Smaller unit cell parameters mean that the bond between the *B* atom in the perovskite and the six coordinating oxygen atoms around it is shorter. And it turns out that Al-O (1.651) has a shorter bond length than Fe-O (1.759),[41] and the bond energy of Al-O is larger, which means that the octahedral structure composed of aluminum atoms and surrounding oxygen atoms is more stable, making the perovskite phase less prone to structural phase change or structural collapse at high temperature or low oxygen partial pressure conditions. However, when $x$ is greater than 0.6, both CPO and $PSF_{1-x}A_xO$ phase cell parameters both become larger, which may be attribute to more $Al^{3+}$ diffuses to the interstitial site of CPO and the PSFAO with the increase of Al reflecting by more impurities of $PrSrAl_3O_7$ showing up with the increase of Al. A similar situation has been reported by Wu.[42]

We next inspected the microscopic morphology of the 60CPO-40PSF$_{1-x}$A$_x$O

membranes after high temperature sintering by scanning electron microscope (SEM). After several attempts at different sintering temperatures, it was finally found that when the sintering temperature was 1475 °C, the surfaces of the obtained composite membranes were dense. As shown in **Fig. S3**, there are no pores and cracks or other impurities on the surface of the membranes, and the grain are tightly bound and clearly demarcated. Further, the CPO and PSF$_{1-x}$A$_x$O phase could be identified by BSEM as shown in **Fig. 3.** The gray grain is the perovskite phase (PSF$_{1-x}$A$_x$O), and the light grain is the fluorite phase (CPO), since the contribution of the backscattered electrons to the SEM signal intensity is proportional to the atomic numbers. It can be clearly observed that the grains of the CPO phase and the PSF$_{1-x}$A$_x$O phase are randomly and uniformly distributed in the membranes, and the average grain sizes of the fluorite and perovskite phases are substantially uniform. Uniform distribution of the two phases helps ensure that each oxygen ion transport channel (is mainly CPO and PSF$_{1-x}$A$_x$O phase) and electron transport channel (is mainly PSF$_{1-x}$A$_x$O phase) are connected. Thus, the oxygen permeation process can be smoothly carried out, and the dual-phase membrane can be predicted to have good oxygen permeation performance. However, when the Al content $x$ reaches 0.8, as shown in figure (g) and (h), dark grain (PrSrAl$_3$O$_7$) appears on the surface of the material, which is consistent with the previous XRD results.

To further study the effect of different Al content on the oxygen permeability of 60CPO-40PSF$_{1-x}$A$_x$O MIEC membranes, we tested the oxygen permeation rates of 60CPO-40PSF$_{1-x}$A$_x$O ($x$ = 0.1, 0.2, 0.3, 0.4, 0.6, 0.8) with different Al content using our homemade device. **Fig. 4** exhibits the oxygen permeation fluxes ($JO_2$) of membranes with a thickness of 0.4 mm with different Al content at the temperature range of 900-1000 °C in air/He condition. The most intuitive conclusion is that the oxygen permeation fluxes through all the 60CPO-40PSF$_{1-x}$A$_x$O ($x$ = 0.1, 0.2, 0.3, 0.4, 0.6, 0.8) dual-phase membranes increase with enhancing temperatures. This is because the increase of temperature provides more internal energy for the diffusion of oxygen ions, so that more oxygen ions can overcome the potential barrier from one side of the membrane to the other side faster.

Of interest to us, the oxygen permeation fluxes of different membranes are not positively or negatively correlated with the content of Al at a certain temperature. For example, at 1000 °C, the $JO_2$ of $PSF_6A_4O$ is the highest (1.12 mL min$^{-1}$ cm$^2$) and $PSF_2A_8O$ has the lowest oxygen permeation fluxes (0.5 mL min$^{-1}$ cm$^2$). In order to more intuitively understand the influence of the change of Al content on the oxygen permeation fluxes of CPO, the broken line graph of $JO_2$ as a function of Al content is plotted in **Fig. 5**. It can be seen clearly that with the increase of aluminum content ($x$), the $JO_2$ of the 60CPO-40PSF$_{1-x}$A$_x$O membranes increases initially. When $x = 0.4$, the $JO_2$ of membranes reaches the maximum value. However, when $x$ continues to increase and is greater than 0.4, $JO_2$ of membranes begins to decrease. When $x$ increases to 0.8, $JO_2$ is minimized. This may be interpreted as the following aspects: First, because the variable valence of $Fe^{3+}/Fe^{4+}$ ions provides better electronic conductivity for the PSF$_{1-x}$A$_x$O phase, excessive substitution of $Fe^{3+}/Fe^{4+}$ by $Al^{3+}$ will lead to a decrease in the electronic conductivity of the material, which leading to the reduce of the oxygen permeability of the 60CPO-40PSF$_{1-x}$A$_x$O membranes. Second, as shown in the previous XRD pattern and SEM graph, an excess of Al doping leading to the formation of impurity phases that enrichment on the surface of the membranes results in a decrease in the oxygen permeability of the 60CPO-40PSF$_{1-x}$A$_x$O composites. When $Al^{3+}$ is substituted a small amount of $Fe^{3+}/Fe^{4+}$, the stoichiometric ratio of oxygen in PSF$_{1-x}$A$_x$O will be lower than that in PSFO, which means that there are more oxygen vacancies in PSFAO, thus improving the oxygen permeability of 60CPO-40PSF$_{1-x}$A$_x$O composites doped with Al ($x \leq 0.4$). Briefly speaking, on the one hand, due to the substitution of $Al^{3+}$ for $Fe^{3+}/Fe^{4+}$, the membrane has more oxygen vacancies, thus higher oxygen permeability. On the other hand, too much substitution of $Al^{3+}$ for $Fe^{3+}/Fe^{4+}$ will lead to the decrease of electronic conductivity of the membrane and the formation of harmful $PrSrAl_3O_7$ phase, which leads to lower oxygen permeability. Therefore, the suitable of Al doping is benefit for the structure stability and oxygen permeability. From the aforementioned analysis results, we can conclude that the optimal doping level for oxygen permeability in 60CPO-40PSF$_{1-x}$A$_x$O membranes is $x$

= 0.4, where the $JO_2$ is 1.12 mL min$^{-1}$ cm$^2$ at 1000 °C and reach the minimum requirements of industrial application (1.0 mL min$^{-1}$ cm$^2$).[43]

**Fig. 6** presents the Arrhenius plots of the oxygen permeation fluxes through 60CPO-40PSF$_{1-x}$A$_x$O membranes with different Al content under an air/He atmosphere. According to **Fig. 6**, the apparent activation energies of 60CPO-40PSF$_{1-x}$A$_x$O ($x$ = 0.1, 0.2, 0.3, 0.4, 0.6, 0.8) in the range of 900-1000 °C were calculated to be 76.36, 67.99, 65.68, 40.39, 51.02 and 51.77 kJ mol$^{-1}$, respectively. Among them, the apparent activation energy of 60CPO-40PSF$_6$A$_4$O is the smallest, which means that the energy required for oxygen ions inside the material to break through the potential barrier and diffuse is the lowest. And pre-exponential factor of 60CPO-40PSF$_6$A$_4$O composition was found to be the largest, indicating that the number of active sites inside 60CPO-40PSF$_6$A$_4$O is more than other 60CPO-40PSF$_{1-x}$A$_x$O membranes. This is consistent with the highest oxygen permeability of 60CPO-40PSF$_6$A$_4$O among the 60CPO-40PSF$_{1-x}$A$_x$O.

**Fig. 7** presents the oxygen permeation fluxes through 60CPO-40PSF$_{1-x}$A$_x$O as a function of time under He/air gradient at 1000 °C. As shown in **Fig. 8**, the oxygen permeation fluxes through all the 60CPO-40PSF$_{1-x}$A$_x$O membranes remain unchanged regardless of the value of the Al content ($x$). And after 50 hours of stable worked, the $JO_2$ of the 60CPO-40PSF$_6$A$_4$O membranes does not decrease and there is no crack on the membranes surface. Other membranes did not have any cracks on their surfaces after 24 hours of stable operation. The results real that 60CPO-40PSF$_{1-x}$A$_x$O membranes can maintain good oxygen permeability and high temperature-resistant stability even when working in a 1000 °C environment for a long time. As previously analyzed, the doping of Al to the $B$-position of the perovskite ($ABO_3$) makes the structure of the oxygen octahedron more stable, so that the structure of the PSF$_{1-x}$A$_x$O phase still has good order in a high temperature environment for a long time.

To further investigate the $CO_2$ stability of 60CPO-40PSF$_{1-x}$A$_x$O composites, the composites powders were exposed in pure $CO_2$ environment for 24 hours at different temperatures (800 °C, 900 °C and 1000 °C). As shown in **Fig. 8,** the XRD patterns of 60CPO-40PSF$_{1-x}$A$_x$O powder after exposing in the atmosphere of carbon dioxide for 24 hours did not change significantly, indicating that the materials did not react with $CO_2$ to generate corresponding carbonate. However, in some previous reports, the formation of carbonate will destroy the phase structure of the membranes and reduce the oxygen permeability and stability of the membranes. The above results signify that our 60CPO-40PSF$_{1-x}$A$_x$O membranes exhibit good structure stability even in pure $CO_2$ atmosphere at 800-1000 °C and can avoid this bad situation. In our previous study, 60Ce$_{0.9}$Pr$_{0.1}$O$_{2-\delta}$-40Pr$_{0.6}$Sr$_{0.4}$Fe$_{0.8}$Al$_{0.2}$O$_{3-\delta}$ dual phase membrane worked stably for 100 hours with $CO_2$ as the sweep gas at 1000 °C. In summary, the 60CPO-40PSF$_{1-x}$A$_x$O membrane has excellent $CO_2$-tolerant stability and it has a wide range of potential application prospects.

**Conclusions**

The Al-containing composites 60wt.%Ce$_{0.9}$Pr$_{0.1}$O$_{2-\delta}$-40wt.%40Pr$_{0.6}$Sr$_{0.4}$Fe$_{1-x}$Al$_x$O$_{3-\delta}$ ($x$ = 0.05, 0.1, 0.2, 0.3, 0.4, 0.6, 0.8, 1.0) was successfully synthesized via a modified Pechini method. The XRD and the refining results show that the doping of Al makes the cell parameters of the PSF$_{1-x}$A$_x$O phase smaller. When $x$ is greater than 0.6, the third phase appeared in the composite, and the cell parameters become larger. The SEM image shows that the composites have excellent chemical compatibility, and the two phases are uniformly mixed and have no obvious pores. Oxygen permeability test showed that the oxygen permeability of 60CPO-40PSF$_{1-x}$A$_x$O membrane increased with the increase of Al content, but when x was greater than 0.6, the oxygen permeability decreased with the increase of aluminum content. The final stability test shows that the Al-doped composites has good $CO_2$-stable reduction-tolerant and high temperature-resistant, so it has potential application prospects in oxy-fuel combustion based on oxygen permeation membrane.

**Acknowledgment**

H. X. Luo acknowledges the financial support by "Hundred Talents Program" of the Sun Yat-Sen University and National Natural Science Foundation of China (21701197).

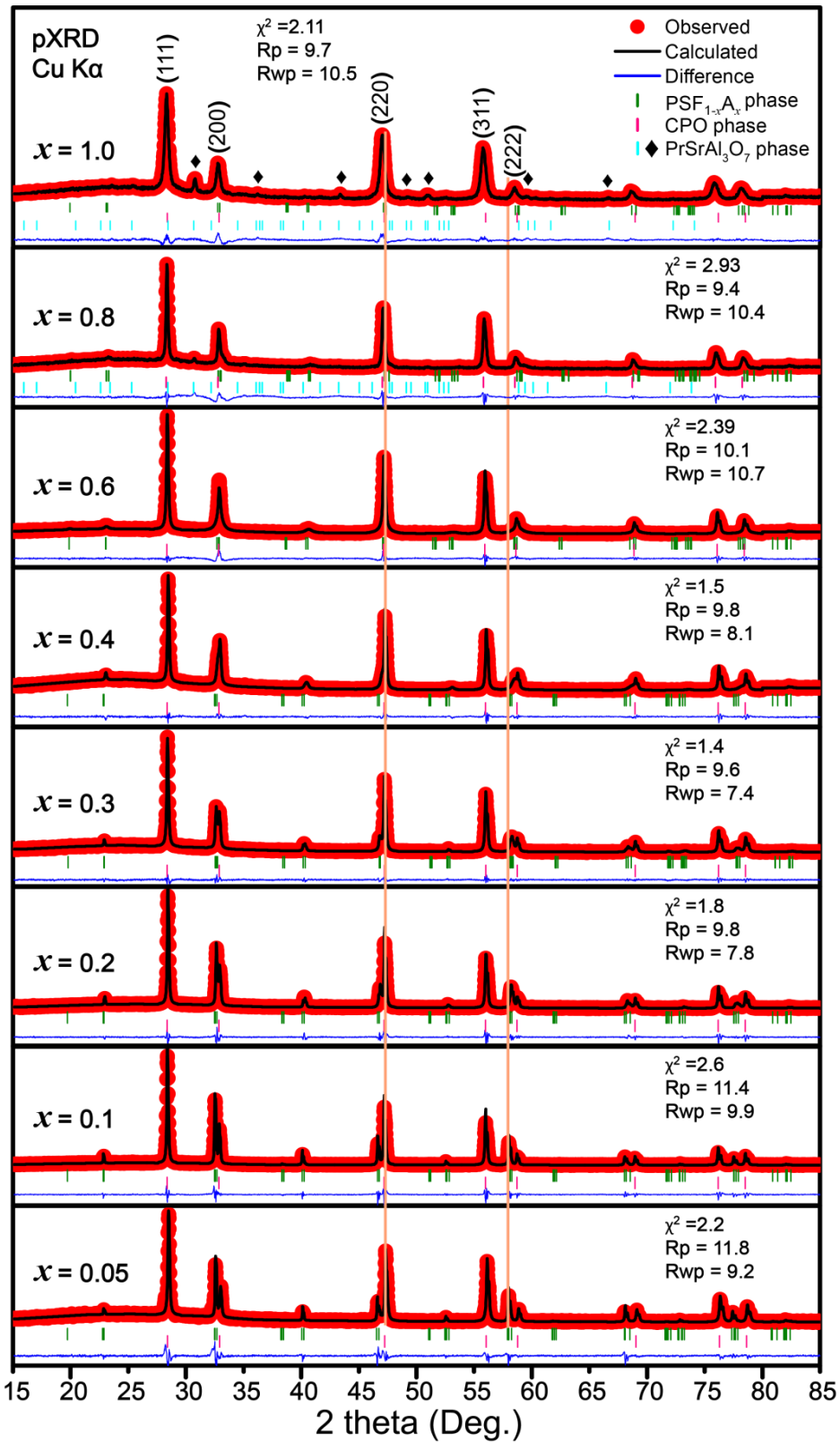

**Fig. 1.** Rietveld refinement XRD patterns of 60CPO-40PSF$_{1-x}$A$_x$O ($x$ = 0.05, 0.1, 0.2, 0.3, 0.4, 0.6, 0.8, and 1.0) powders after calcined at 950 °C for 10 h in air.

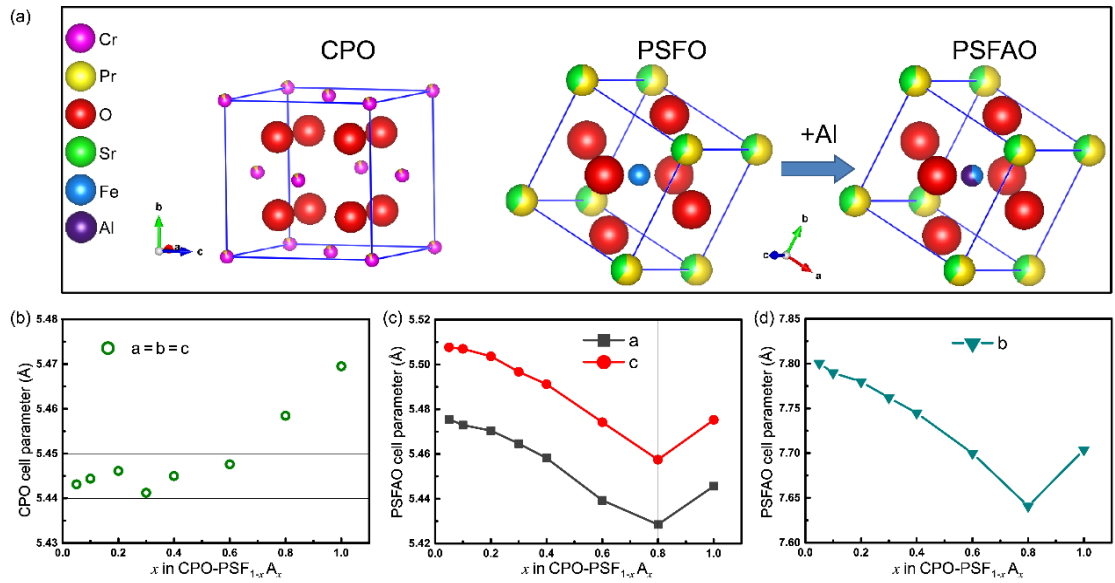

**Fig. 2.** Crystal structure characterization of 60CPO-40PSF$_{1-x}$A$_x$O: (a) the CPO phase crystal structure on the left, the PSF$_{1-x}$A$_x$O phase crystal structure on the right; CPO cell parameter as a function of Al content for (b); PSF$_{1-x}$A$_x$O cell parameter as a function of Al content for (c, d).

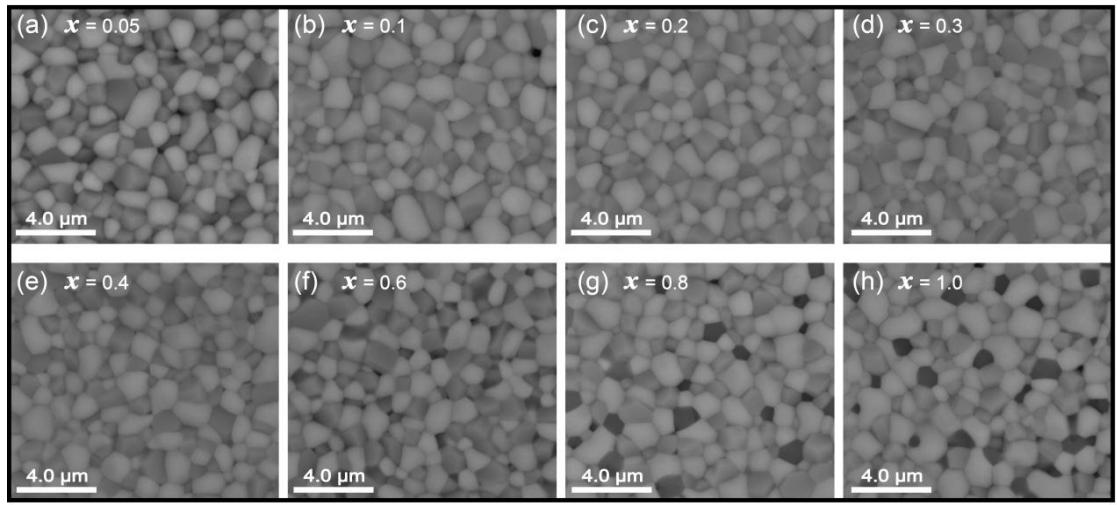

**Fig. 3**. BSEM micrographs of plane view of the surfaces of 60CPO-40PSF$_{1-x}$A$_x$O ($x$ = 0.05, 0.1, 0.2, 0.3, 0.4, 0.6, 0.8, and 1.0) composite membranes after sintering at 1475 °C for 5 h. In BSEM, the gray grains represent the PSF$_{1-x}$A$_x$O grains; the light ones represent the CPO grains, and the dark ones represent the PrSrAl$_3$O$_7$ grains.

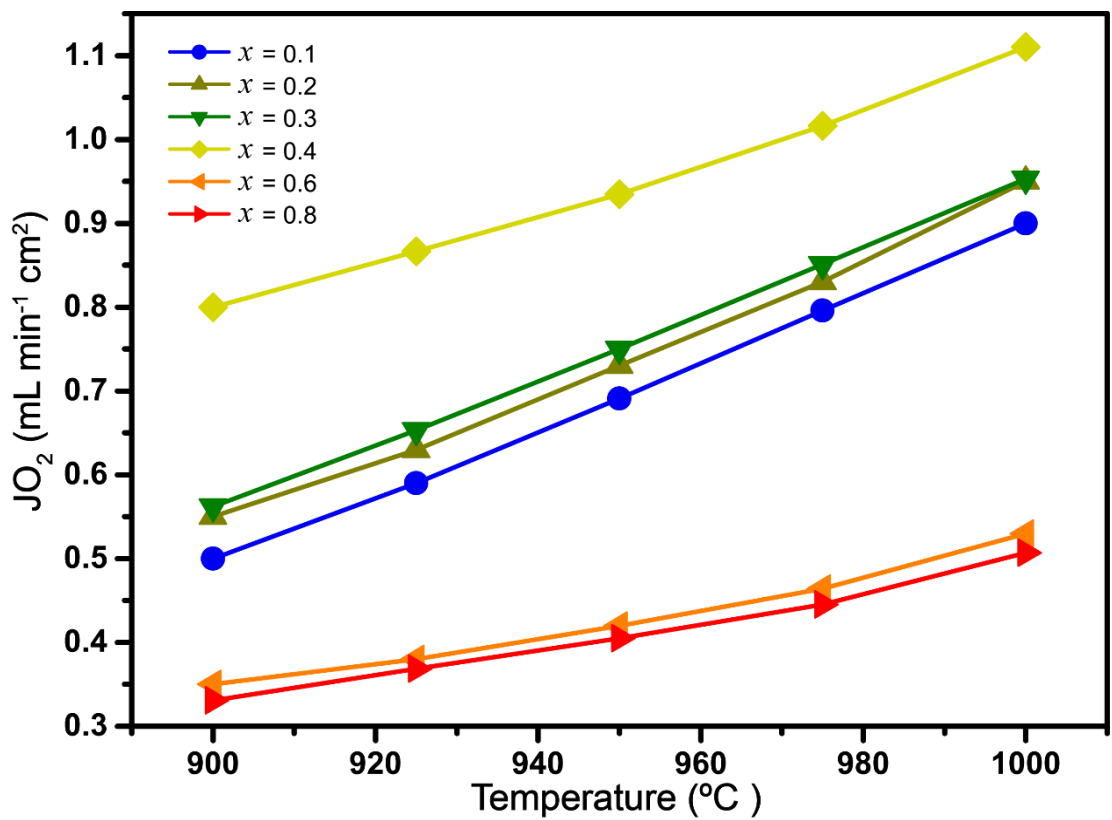

**Fig. 4**. Oxygen permeation fluxes through the 60CPO-40PSF$_{1-x}$A$_x$O ($x$ = 0.1, 0.2, 0.3, 0.4, 0.6, 0.8) composite membranes.

*Condition: 100 mL min$^{-1}$ air as the feed gas, 49 mL min$^{-1}$ He, 1 mL min$^{-1}$ Ne as an internal standard gas. Membrane thickness: 0.4 mm.*

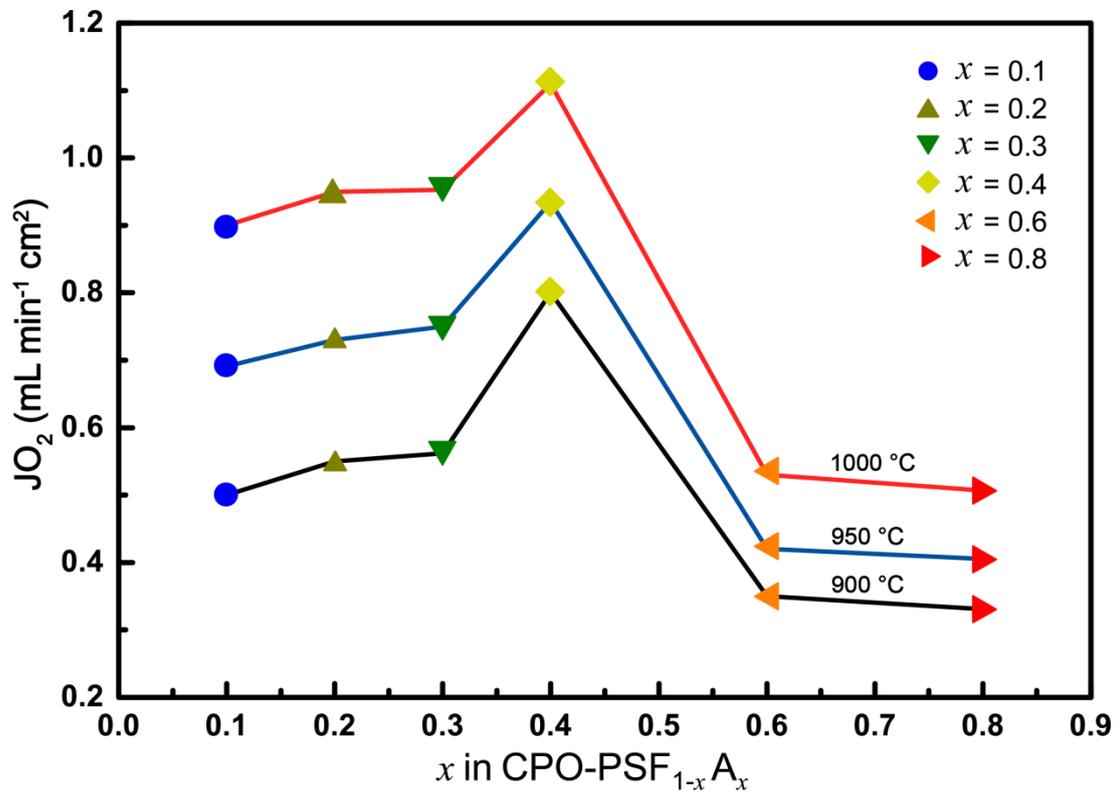

**Fig. 5**. Oxygen permeation fluxes through the 60CPO-40PSF$_{1-x}$A$_x$O ($x$ = 0.1, 0.2, 0.3, 0.4, 0.6, 0.8) composite membranes at 900 °C, 950 °C, 1000 °C.

*Condition: 100 mL min$^{-1}$ air as the feed gas, 49 mL min$^{-1}$ He, 1 mL min$^{-1}$ Ne as an internal standard gas. Membrane thickness: 0.4 mm.*

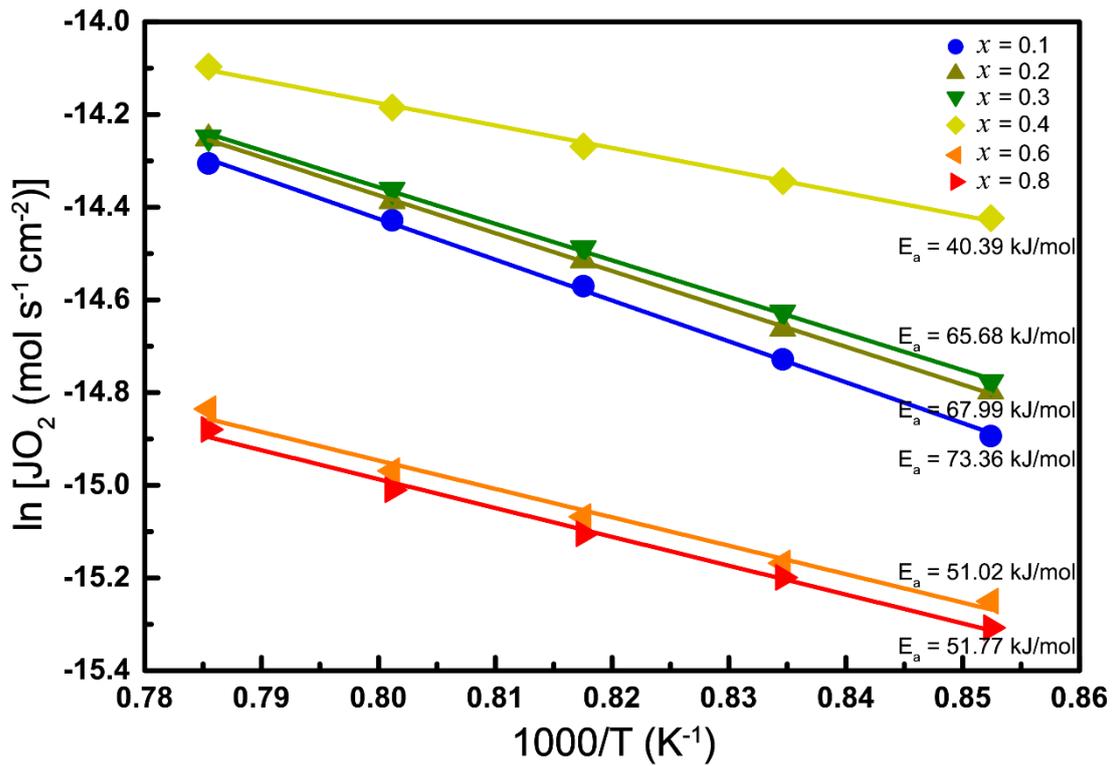

**Fig. 6**. Arrhenius plot of oxygen permeation fluxes for 60CPO-40PSF$_{1-x}$A$_x$O ($x$ = 0.1, 0.2, 0.3, 0.4, 0.6, 0.8) composite membranes.

*Condition: 100 mL min$^{-1}$ air as the feed gas, 49 mL min$^{-1}$ He, 1 mL min$^{-1}$ Ne as an internal standard gas. Membrane thickness: 0.4 mm.*

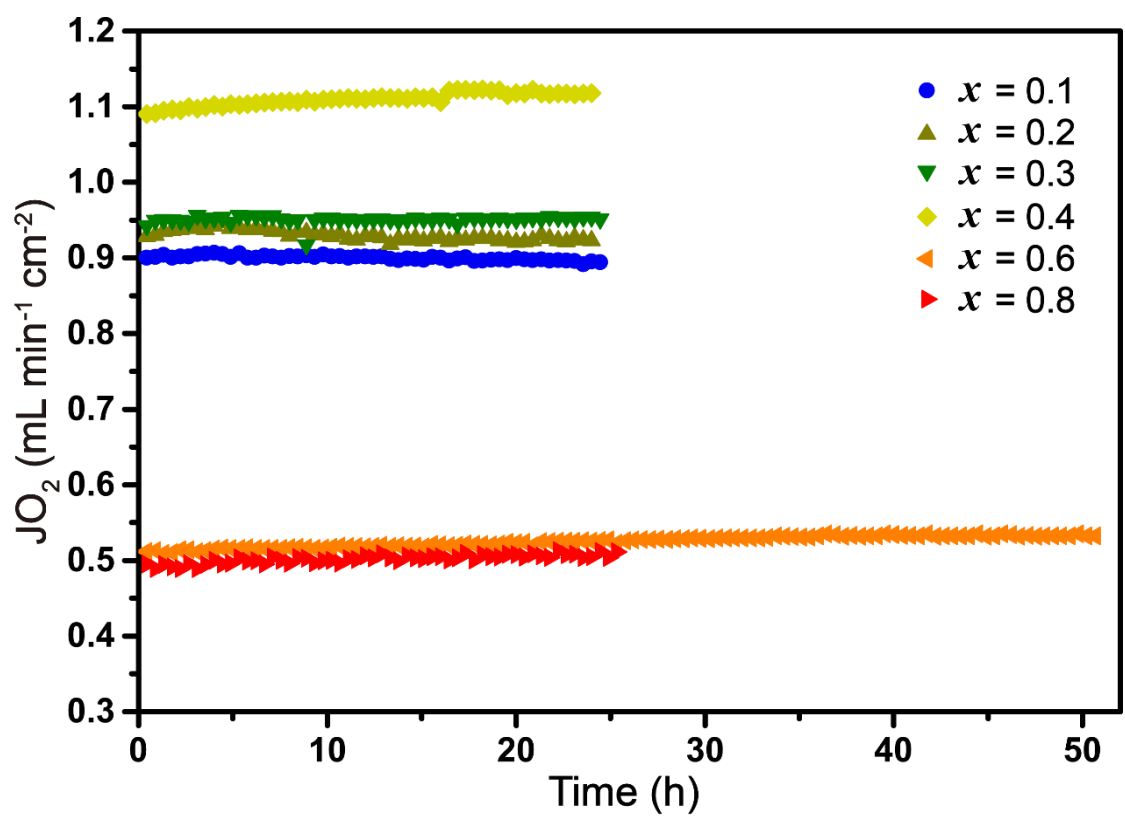

**Fig. 7**. Oxygen permeation fluxes of 60CPO-40PSF$_{1-x}$A$_x$O membranes as a function of time using pure He as sweep gas at 1000 °C.

*Condition: 100 mL min$^{-1}$ air as the feed gas, 49 mL min$^{-1}$ He, 1 mL min$^{-1}$ Ne as an internal standard gas. Membrane thickness: 0.4 mm.*

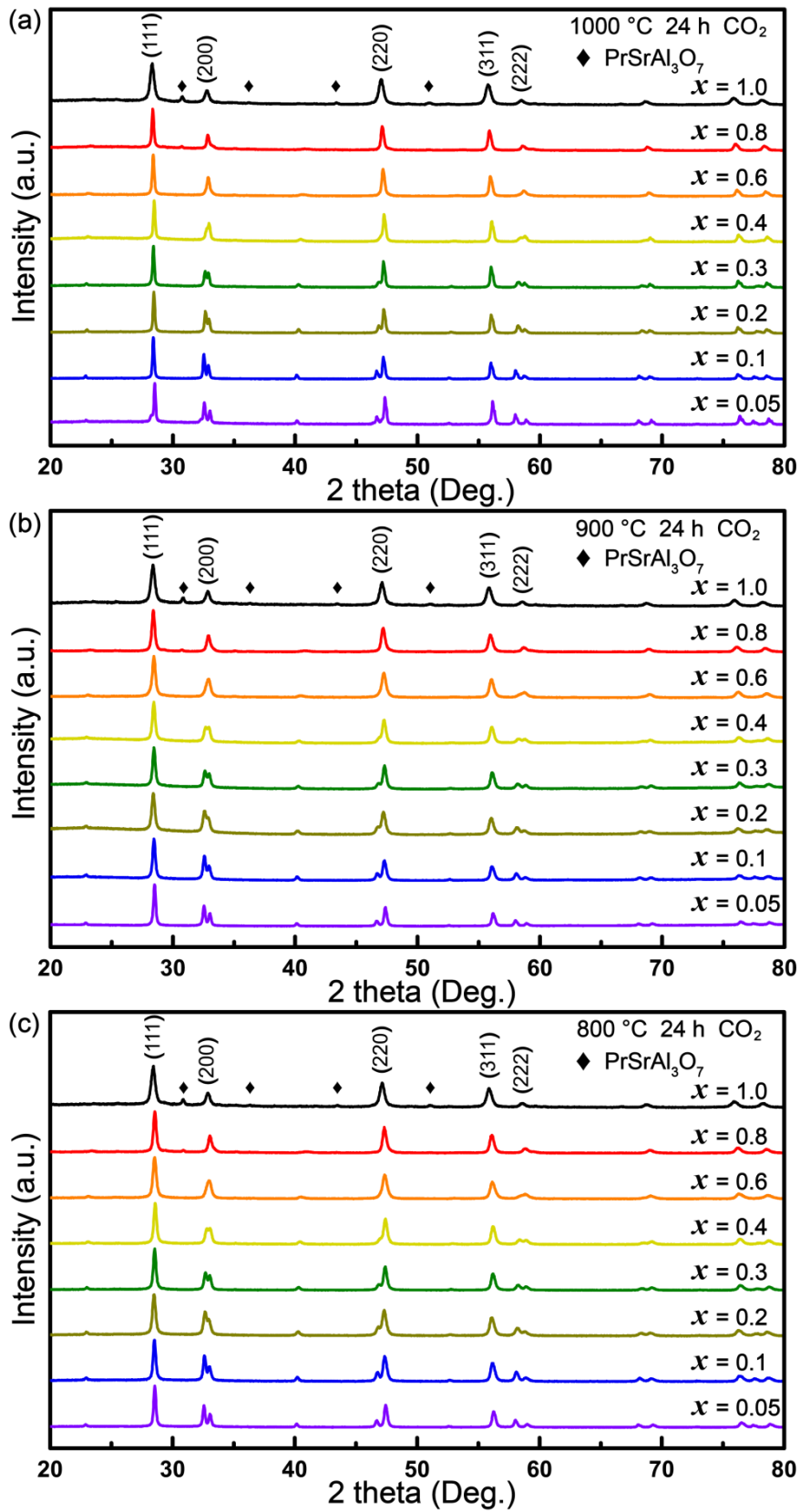

**Fig. 8**. XRD patterns of 60CPO-40PSF$_{1-x}$A$_x$O powders calcined at 800 °C, 900 °C and 1000 °C for 24 h under CO$_2$ atmosphere.

# Supporting information

# Effects of Al content on the oxygen permeability through dual-phase membrane 60Ce$_{0.9}$Pr$_{0.1}$O$_{2-\delta}$-40Pr$_{0.6}$Sr$_{0.4}$Fe$_{1-x}$Al$_x$O$_{3-\delta}$

Lei Shi[a], Shu Wang[a], Tianni Lu[b], Yuan He[a], Dong Yan[a], Qi Lan[a], Zhiang Xie[a], Haoqi Wang[a], Mebrouka Boubeche[a], Huixia Luo[a]*

[a]*School of Material Science and Engineering and Key Lab Polymer Composite & Functional Materials, Sun Yat-Sen University, No. 135, Xingang Xi Road, Guangzhou, 510275, P. R. China*

[b]*School of Materials Sciences and Engineering, Shenyang Aerospace Unversity, Shenyang, 110136, P. R. China*

*Corresponding author/authors complete details (Telephone; E-mail:) (+0086)-2039386124

luohx7@mail.sysu.edu.cn

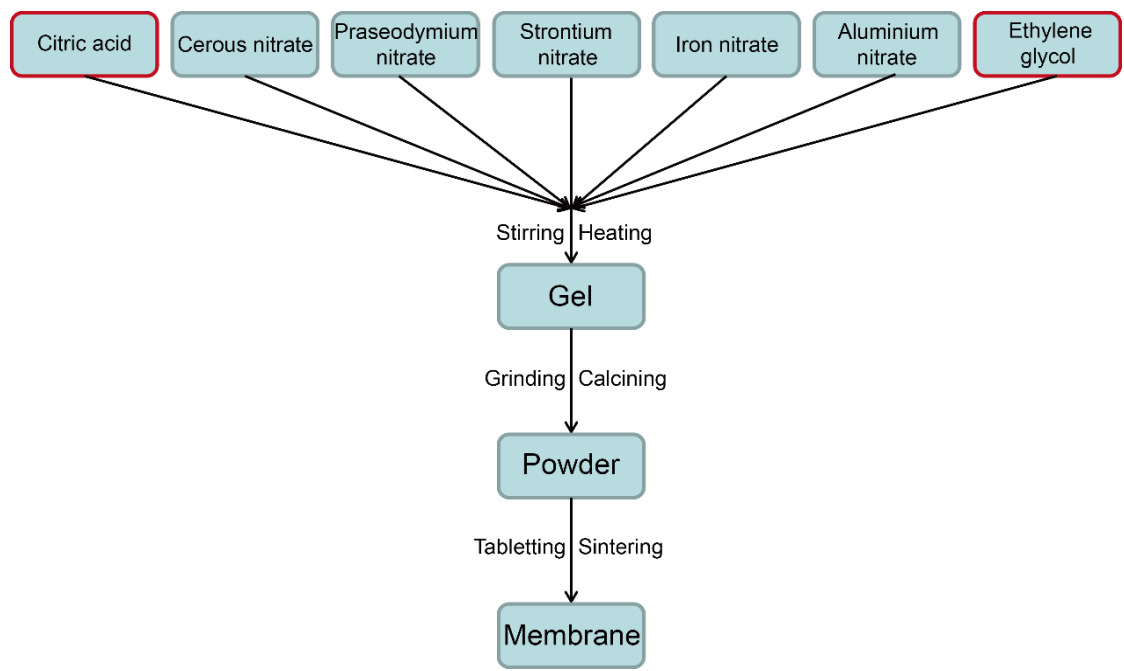

**Fig. S1.** The flowchart of preparation process of 60CPO-40PSF$_{1-x}$A$_x$O via a modified Pechini method.

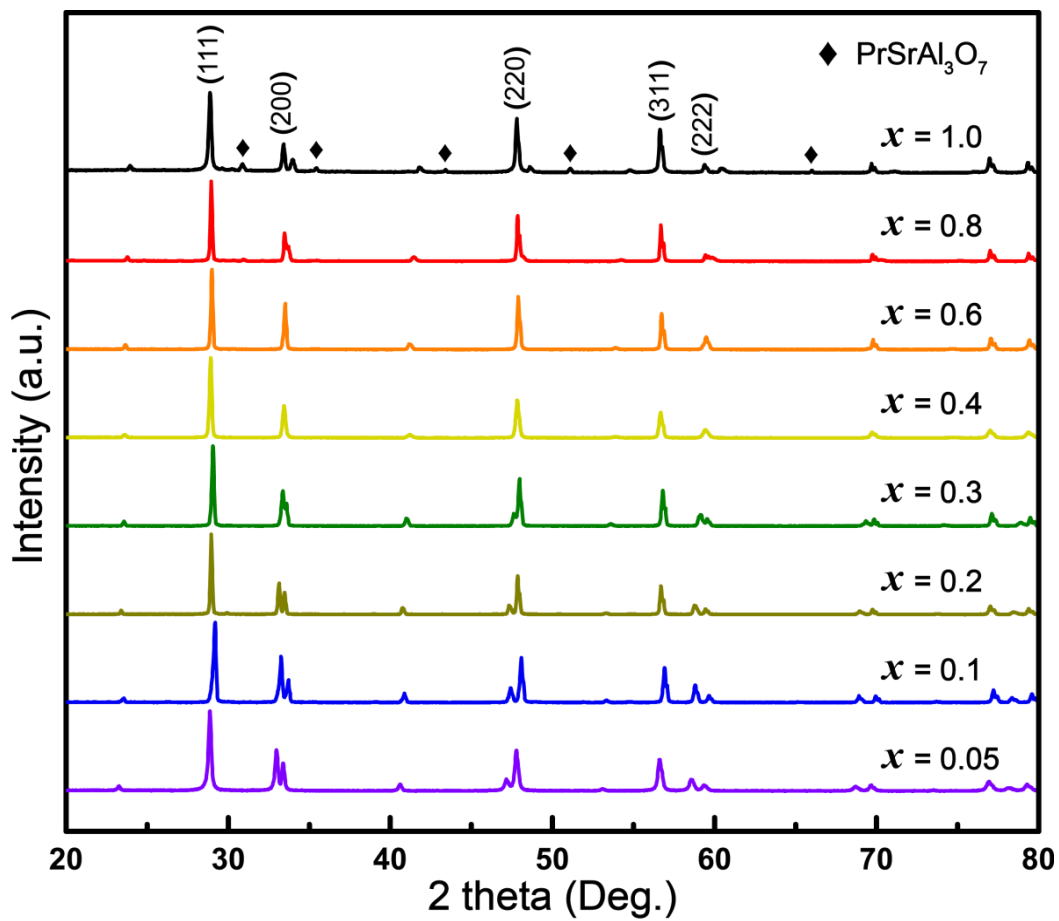

**Fig. S2** XRD patterns of 60CPO-40PSF$_{1-x}$A$_x$O composite membranes after sintering at 1475 °C for 5 h.

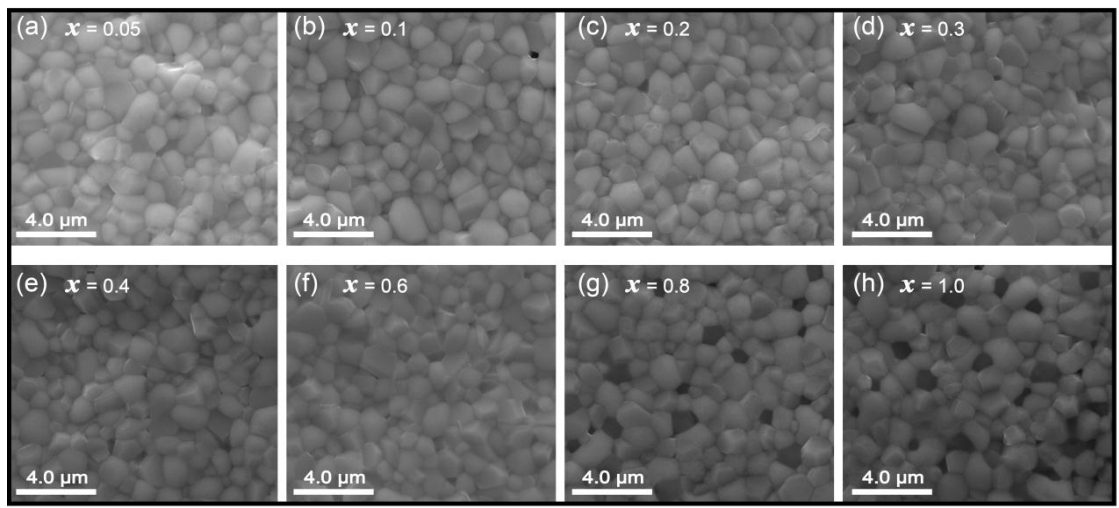

**Fig. S3** SEM micrographs of plane view of the surfaces of 60CPO-40PSF$_{1-x}$A$_x$O ($x$ = 0.05, 0.1, 0.2, 0.3, 0.4, 0.6, 0.8, and 1.0) composite membranes after sintering at 1475 °C for 5 h.